\documentstyle[12pt, A4]{article}
\thicklines                     
\begin{document}
\begin{titlepage}
\begin{centering}
\vspace{2cm}
{\Large\bf Coordinate-space holographic projection of fields }\\
\vspace{0.5cm}
{\Large\bf and an application to massive vector fields}\\
\vspace{2cm}
W.~S.~l'Yi\footnote{E-mail address: {\tt wslyi@cbucc.chungbuk.ac.kr}}
\vspace{0.5cm}

{\em Department of Physics}\\
{\em Chungbuk National University}\\
{\em Cheongju, Chungbuk 361-763, Korea}\\
\vspace{1cm}
\begin{abstract}
\noindent  General properties of coordinate-space holographic projections 
of fields in AdS/CFT correspondence, which respect the Ward identity, 
are investigated.  
To show the usefulness of this methodology it is  applied
to the computation of correlators of massive gauge fields. 
\end{abstract}
\end{centering} 
\vfill
\noindent HEP-CNU-9809\\
hep-th/9808051\\
Aug.~1998\\
Revised on Jan.~8, 1999.
\end{titlepage}

\setcounter{footnote}{0}
\section{Introduction}                        
\label{Sect1}
The fascinating proposal of Maldacena\cite{Maldacena} that large $N$ limit of a 
certain conformal field theory (CFT) in a $d$-dimensional space can be 
considered as a boundary theory of $d+1$-dimensional anti-de Sitter space (AdS) 
with a compact extra space opened a novel way of understanding various facets of
supersymmetric gauge theories. 
The baryonic vertex, that is the finite energy configurations with
$N$ external quarks, was shown to exist in this context\cite{Witten_baryon}.
Furthermore it was exected that the quark-confining potential of QCD could be 
investigated in this way\cite{applications}.  This AdS/CFT correspondence was
also used to the topic of black hole entropy\cite{black_hole}.

The underlining principle behind this AdS/CFT correspondence was surveyed  
by Gubser {\it et al.}\cite{Gubser} and Witten\cite{Witten_holography}. 
The new insight was that the symmetry  
group $SO(2,d)$ which acts on $AdS_{d+1}$ acts conformally in the boundary 
of $AdS_{d+1}$ which is in fact the Minkowski space.  The holographic properties of the 
Maldacena's proposal were investigated, and various correlation 
functions of conformal
fields were obtained.  The calculations produced the correct conformal
dimensions. 
Motivated by this many authors walked along the path of that 
understanding\cite{Witten's_path, Ya_1, Ya_2, Muck}.  
Since gauge fields can obtain masses from the excitation of the extra compact 
manifold of the AdS/CFT correspondence it is quite natural to
investigate the holographic projection of massive vector fields.  
In fact M\"uck and Viswanathan challenged this problem\cite{Muck} with  
deep understanding.  

But to get the correlators which preserve the Ward identity one usually works on 
the momentum-space with given specific solutions of classical equations of motion.  
It is sometimes quite cumbersome and makes it difficult to get insights on what 
is going on in the coordinate-space.  
One of the purpose of this paper is to examine the holographic 
projections of general fields in AdS/CFT correspondence in the coordinate-space.  
To preserve the necessary Ward identity we are 
careful to approach the boundary of AdS uniformly for all fields.
This result is applied to massive vector fields in a different way than
\cite{Muck}.  The final result of cause agrees with the known one.

The next section is devoted to investigate the general properties of 
coordinate-space holographic projection of fields. 
The final section is an application.
\section{Holographic projections of fields in coordinate-space}
To depict the essencial point behind current paper more clearly, 
we  briefly introduce the
idea of the holographic projection  of Witten\cite{Witten_holography}.
It is known that the boundary of $AdS_{d+1}$  
is the compactified version of Minkowski space ${\cal M}^*_d.$  
For simplicity we assume the Euclidean field theory.
There are many equivalent
choices of the metrices of $AdS_{d+1}.$
In this paper we represent $AdS_{d+1}$ as the Lobachevsky space
${\bf R}^{1+d}_+ = \{(x_0,{\bf x})\in {\bf R}^{1+d}\; | \;x_0>0\}$
with the metric      
\begin{equation}
ds^2={1\over x_0^2}dx^\mu dx^\mu. \label{metric}
\end{equation}  
In this representation distances between points in the $x_0 \to \infty$ region 
vanish. This means that the boundary of $AdS_{d+1}$ is in fact
${\bf S}^d,$ the $d-$dimensional sphere. On the other hand  
the metric (\ref{metric}) diverges as $x_0 \to 0,$ which also means that every 
point in the holographically projected space ${\cal M}^*_d$ is infinitely 
seperated.  To resolve this infinite scale problem one needs
a rescaling function $\mu(x_0)$ which necessarily diverges as $x_0 \to 0.$ 
Using this, distances $\ell_{AdS}$ in $AdS_{d+1}$ can be 
holographically projected to
\begin{equation}
\ell_{{\cal M}^*} = \lim_{x_0 \to 0} {\ell_{AdS}\over\mu},
\end{equation}
giving finite results.  But this rescaling function  $\mu(x_0)$ is
not unique.  When one rescale the rescaling function
\begin{equation}
\mu(x_0) \to e^{-\sigma(x_0)}\mu(x_0)\label{rescale},
\end{equation}
the scale in ${\cal M}^*_d$ changes as
\begin{equation} 
\ell_{{\cal M}^*} \to e^{\sigma} \ell_{{\cal M}^*},   \label{length}
\end{equation}                                   
showing that ${\cal M}^*_d$ has possibly some conformal structure.

Now consider a dynamical field $\phi.$  
Asymptotically this field either vanishes or diverges on the boundary.
In usual field theory the divergent component is discarted.
But in AdS/CFT correspondence one may rescale the divergent 
component $\phi_{AdS}$ with a proper $\lambda$ by
\begin{equation}
\phi_{\cal M^*} = \lim_{x_0 \to 0}{\phi_{AdS}\over \mu^{\lambda}}.
\end{equation}  
Under the change (\ref{rescale}) of rescaling function the field 
$\phi_{{\cal M}^*}$ transforms
as
\begin{equation}
\phi_{\cal M^*} \to e^{\lambda\sigma} \phi_{\cal M^*}.  
\end{equation}
This shows that $\phi_{\cal M^*}$ is a conformal density of the 
length dimension
$\lambda.$  

But in practice one should be careful to approach fields $\phi(x_0,{\bf x})$
and $\partial_0\phi(x_0,{\bf x})$ to the $x_0\to 0$ boundary at the same rate.
If it is not true, the Ward identity will be destroyed\cite{Witten's_path}.  
For this one 
first approach the $x_0\to \epsilon$ boundary instead of the real one.
After the computation of physical quantities in this
$\epsilon$-boundary, the $\epsilon\to 0$ limit is taken to get 
the desired results.  
This process is usually done in the momentum-space, and it is quite cumbersome.
We consider this process in the coordinate space which would be more 
intuitive and practical.
                                                        
Consider an action which is a functional of both $\phi(x_0, {\bf x})$
and $\partial_\mu\phi(x_0, {\bf x}).$  
As long as there is no possibility of confusion we suppress the subscripts
such as $AdS$ and ${\cal M^*.}$
For the holographic projection of
fields we need to consider $\phi(x_0, {\bf x})$ and 
$\partial_0\phi(x_0, {\bf x}).$  
First we define $\epsilon$-boundary field $\phi_{\epsilon}({\bf x})$ by
\begin{equation}
\phi_{\epsilon}({\bf x}) = \phi(\epsilon,{\bf x}).
\end{equation}
Then this boundary field is related to the bulk field $\phi(x_0, {\bf x})$                                           
in the following way
\begin{equation}
\phi(x_0, {\bf x}) = \int d^d x'\; G_\epsilon (x_0, {\bf x}-{\bf x}')
                    \phi_{\epsilon}({\bf x}'). 
\label{bulk_phi}
\end{equation}
Here the boundary-to-bulk Green's function $G_\epsilon (x_0, {\bf x}-{\bf x}')$ 
has following property
\begin{equation}
\lim_{x_0\to \epsilon}G_\epsilon (x_0, {\bf x}-{\bf x}') 
        =\delta^{(d)} ({\bf x}-{\bf x}').
\end{equation}
Since this $\epsilon$-limit procedure is independent of ${\bf x},$
we supress it for notational convenience.   It makes us not to 
worry about the integration with respect to ${\bf x}.$

Differentiating (\ref{bulk_phi}) with respect to $x_0$ and using the above 
mentioned convention we have
\begin{equation}
\partial_0 \phi(x_0) = \partial_0 G_\epsilon (x_0)
                           \phi_{\epsilon}.
\label{partial_phi}
\end{equation}
Now suppose that $\phi(x_0)$ diverges as $x_0^{-\lambda}$ as $x_0\to 0.$
In this case the holographically projected field $\phi_h(\epsilon)$ to
the $\epsilon$-boundary is defined by
\begin{equation}
\phi_h(\epsilon) = \epsilon^\lambda \phi(\epsilon)
            =\phi_h(0)+{\cal O}(\epsilon),
\label{phi_h}
\end{equation}
where ${\cal O}(\epsilon)$ is a function of $\epsilon$ which
vanishes as $\epsilon\to0.$
For this field we have
\begin{equation}
G_\epsilon(x_0)\phi_h (\epsilon) = 
   \left({\epsilon \over x_0} \right)^\lambda \phi_h(x_0).
\end{equation}
Plugging this into (\ref{partial_phi}) we are able to write $\partial_0\phi(x_0)$ 
at the $\epsilon$-boundary in terms of the $\epsilon$-boundary field 
$\phi_\epsilon$ in the following way
\begin{equation}
\left.\partial_0\phi(x_0)\right|_{x_0=\epsilon} = 
      \left({\partial_\epsilon \phi_h(\epsilon) \over \phi_h(\epsilon) }
      - {\lambda\over\epsilon}\right) \phi_\epsilon.\label{partial_phi_}
\end{equation}                                                             
To compare the degree of divergence between two terms in the parenthesis 
we use (\ref{phi_h}),
\begin{equation} 
\left.\partial_0\phi(x_0)\right|_{x_0=\epsilon} 
  =\left\{ {\partial_\epsilon {\cal O}(\epsilon) \over \phi_h(0) }
    \left(1 - { {\cal O}(\epsilon) \over \phi_h(0)} +...\right)
      - {\lambda\over\epsilon}\right\}\phi_\epsilon.
\end{equation}
This shows that the last term,
when compared to the others, is a divergent one which
is related to the delta-function contact term 
in the computation of correlators.  So we drop it.  This process of 
regularization of the infinity is the coordinate-space version of the one
discussed in the appendix of {\it Ref.}\cite{Witten's_path}.    

The final result of the holographic projection
of $\partial_0 \phi(x_0,{\bf x})$ is
\begin{equation}
\left.\partial_0 \phi(x_0,{\bf x})\right|_{x_0=\epsilon}
   \to \epsilon^{-\lambda} \partial_\epsilon \phi_h(\epsilon,{\bf x}).
\end{equation}
It can be compared to the holographic projection of $\phi(x_0,{\bf x}),$
\begin{equation}
\left.\phi(x_0,{\bf x})\right|_{x_0=\epsilon}
   \to \epsilon^{-\lambda} \phi_h(\epsilon,{\bf x}). 
\end{equation}
This method of holographic projection of fields in AdS/CFT correspondence
is more intuitive and practical than the one which is done in the momentum-space.
As an example, the holographic projection of 
$\phi(x_0,{\bf x})\partial_0\phi(x_0,{\bf x})$ can be written as
\begin{equation}
\left.\phi(x_0,{\bf x})\partial_0\phi(x_0,{\bf x})\right|_{x_0=\epsilon}
    \to \epsilon^{-2\lambda}\partial_\epsilon {1\over 2}  
     \phi_h (\epsilon,{\bf x})^2.
\label{example}
\end{equation}
We apply this idea to massive vector fields in the next section.
\section{Holographic projection of massive gauge fields}
\label{Sect2}

According to Maldacena the generating functional 
$\langle \exp \int d^dx A_i({\bf x}) J_i({\bf x}) \rangle_{CFT}$ of the
correlation functions 
$\langle J_{i_1}({\bf x}_1)J_{i_2}({\bf x}_2)\dots J_{i_n}({\bf x}_n) \rangle$
can be constructed from the following proposal
\begin{equation}
\langle \exp \int d^dx A_i({\bf x}) J_i({\bf x}) \rangle_{CFT}\label{proposal}
= {\cal Z}_{AdS_{d+1}}[{\cal A}_\mu],
\end{equation}
where ${\cal Z}_{AdS_{d+1}}[{\cal A}_\mu]$ is a supergravity partition 
function on $AdS_{d+1}$ computed with the condition that ${\cal A}_\mu$ projects
to $A_i$ at the boundary of $AdS_{d+1}$ which is in fact the compactified
$d-$dimensional Minkowski space.  When ones use the classical 
approximation this partition function can be written as
\begin{equation}
{\cal Z}_{AdS_{d+1}}[{\cal A}_\mu] \simeq \exp(-I[A_i]),
\label{proposal_2}
\end{equation}
where $I[A_i]$ is the projected classical action.  We start from this 
classical approximation to compute the two-point correlation function of
the conformal current $J_i$ for massive vector fields.

The action for the massive gauge field in $AdS_{d+1}$ space is
\begin{equation}
I = \int d^{d+1}x \sqrt g\;\left( {1\over 4}{\cal F}_{\mu\nu}{\cal F}^{\mu\nu} 
  + {1\over 2} m^2 {\cal A}_\mu {\cal A}^\mu \right),
\end{equation}
where ${\cal F}_{\mu\nu}=\partial_\mu {\cal A}_\nu - \partial_\nu {\cal A}_\mu$ 
is the usual field strength tensor of a vector field ${\cal A}_\mu.$
The Euler-Lagrange equation of motion which is derived from this action is
\begin{equation}
\nabla^\mu {\cal F}_{\mu\nu}-m^2 {\cal A}_\nu=0. \label{equation_of_motion}
\end{equation}         
To get the holographic projection of the action we use the classical
equation of motion, producing the following action 
\begin{equation}
I = \lim_{\epsilon\to 0}{1\over 2}\int_{x_0=\epsilon} d^d x \sqrt{g} 
  {\cal A}_{\mu} {\cal F}^{\mu 0}.
\end{equation}
As it is pointed out by M\"uck and Viswanathan\cite{Muck} it is useful to define 
fields with Lorentz indices by
\begin{equation}
{A}_a=e^\mu_a {\cal A}_\mu,
\end{equation}
where $e^\mu_a=x_0\delta^\mu_a,$ $(a=0,1 \dots d)$ is the vielbein of $AdS_{d+1}.$ 
Using $A_0(x_0,{\bf x})$ and $A_i(x_0,{\bf x})$ the 
classical action is 
\begin{equation}
I= \lim_{\epsilon\to 0}{1\over 2}\int d^dx \epsilon^{-d}
\left[- {\epsilon\over 2}\partial_0 A_i^2 +
  A_i^2 + \epsilon A_i \partial_i A_0\right].
\label{final_action}
\end{equation}
To express this action in terms of the holographically projected fields
we need to solve the equation of motion
(\ref{equation_of_motion}).  It is not difficult to show that 
for ${\cal A}_0(x_0,{\bf x})$ it is 
\begin{equation}\label{eom}
[\;x_0^2\partial_\mu\partial_\mu + (1-d)x_0\partial_0 - (m^2 - d+1)\;]
 {\cal A}_0 =0.
\end{equation} 
As $x_0 \to 0,$ ${\cal A}_0$ behaves as $x_0^{-\lambda},$ where $\lambda$ is the 
larger root of the following quadratic equation,
\begin{equation}
\lambda(\lambda+d) = m^2-d +1.  \label{lambda}
\end{equation}
The solution of Eq.(\ref{eom}) is well known from the 
scalar field theory\cite{Ya_1, Ya_2, scalar_theory}, and is given by
\begin{equation}
{\cal A}_0 (x_0, {\bf x}) = x_0^{d/2}\int {d^d k \over (2\pi)^d} 
    e^{-i{\bf k}\cdot{\bf x}}
    a_0({\bf k})K_\nu(|{\bf k}|x_0), \label{cal_A_0}
\end{equation}
where 
\begin{equation}
\nu=\lambda+{d\over 2},
\end{equation}
and $K_\nu$ is the modified Bessel function of the third kind which satisfies
\begin{equation}
\left[\,\xi^2{d^2 \over d\xi^2} + \xi{d \over d\xi} 
    -(\nu^2 +\xi^2)\,\right]K_\nu(\xi) = 0.
\end{equation}

The equation of motion (\ref{equation_of_motion}) for 
$A_i,$ $i=1 \dots d,$ reduces to
\begin{equation}
[\,x_0^2\partial_\mu\partial_\mu + (1-d)x_0\partial_0 - (m^2 - d+1)\,]A_i 
    = 2x_0 \partial_i A_0.\label{eom2}
\end{equation}
The exact solution of this equation is considered in \cite{Muck}.  But since
it is sufficient to know the $x_0\to 0$ behaviour of the solution, we follow 
a different path.  First we decompose $A_i(x_0, {\bf x})$ into 
\begin{equation}
A_i (x_0, {\bf x}) = A_i^{(g)} (x_0, {\bf x}) + A_i^{(p)} (x_0, {\bf x}),
\end{equation}
where 
\begin{equation}
A_i^{(g)} (x_0, {\bf x}) =  x_0^{d/2}\int {d^d k \over (2\pi)^d} 
    e^{-i{\bf k}\cdot{\bf x}}
    a_i({\bf k})K_\nu(|{\bf k}|x_0) 
\end{equation}
is the general solution of (\ref{eom2}) when the right hand side vanishes, and
$A_i^{(p)} (x_0, {\bf x})$ is the particular solution which is assumed to have 
following form
\begin{equation}                      
A_i^{(p)} (x_0, {\bf x}) =-2i x_0^{d/2}\int {d^d k \over (2\pi)^d} 
    e^{-i{\bf k}\cdot{\bf x}}
    a_0({\bf k}) {k_i \over |{\bf k}|^2} H(|{\bf k}|x_0). \label{A_p}
\end{equation}                          
Here $H(|{\bf k}|x_0)$ is an unknown function which must be determined.
Both $a_i({\bf k})$ and $a_0({\bf k})$ are independent unknown coefficients
which are related to the projected fields $A_i({\bf x}).$ 
For simplicity we denote $\xi=|{\bf k}| x_0.$  From (\ref{eom2}) it is easy 
to show that $H(\xi)$ satisfies
\begin{equation}
[\,\xi^2{d^2 \over d\xi^2} + \xi{d \over d\xi} 
    -(\nu^2 +\xi^2)\,]H = \xi^2 K_\nu.
\end{equation} 
This ordinary differential equation can be solved by making use of the
the consistency condition $\nabla^\mu {\cal A}_\mu = 0$ of the massive gauge 
field,
\begin{equation}
\partial_i A_i + \partial_0 A_0 - {d \over x_0} A_0=0. 
\end{equation}
For further simplification we separate this into the following two relations,
\begin{eqnarray}
\partial_i A_i^{(g)} &=&0, \label{sub_g} \\
\partial_i A_i^{(p)} + \partial_0 A_0 - {d \over x_0} A_0 &=&0.\label{sub_p}
\end{eqnarray}
The first constraint (\ref{sub_g}) can easily be solved producing
\begin{equation}
k_{i}a_i({\bf k}) = 0.  \label{k_ia_i=0}
\end{equation}
When one substitutes (\ref{cal_A_0}) and (\ref{A_p}) into (\ref{sub_p})
one determines $H(|{\bf k}|x_0)$ in the following way
\begin{equation}
H(\xi)={1\over 2}[\xi {d K_\nu \over d\xi} +(1-{d\over 2})K_\nu].
\end{equation}

Using the expansion                          
\begin{equation}
K_\nu(\xi) = {1\over 2}                
         \left[\;\Gamma(\nu)\left({\xi\over 2}\right)^{-\nu}\!\!\!\!\! 
          +\Gamma(-\nu)\left({\xi\over 2}\right)^{\nu}+\dots\;\right],
\end{equation}
we have 
\begin{equation}
A_i^{(p)} (x_0, {\bf x}) =-2i x_0^{d/2}\int {d^d k \over (2\pi)^d} 
    e^{-i{\bf k}\cdot{\bf x}}
    a_0({\bf k}) {k_i \over |{\bf k}|^2}
\left[ {c\over 2}\Gamma(\nu)\left({\xi\over 2}\right)^{-\nu}\!\!\!\!\! + 
       {\bar{c}\over 2}\Gamma(-\nu)
     \left({\xi\over 2}\right)^{\nu} +\dots\right],\label{a_i(p)_expansion_new}
\end{equation} 
where
\begin{equation}
      c = {1\over 2}(1 - {d \over 2} - \nu ),\;\;\;
\bar{c} = {1\over 2}(1 - {d \over 2} + \nu ). \label{c}
\end{equation}
On the other hand  the series expansion of $A_i^{(g)}(x_0, {\bf x})$ 
is
\begin{equation}
A_i^{(g)}(x_0, {\bf x}) 
 = x_0^{d/2}\int {d^d k \over (2\pi)^d}
  e^{-i{\bf k}\cdot{\bf x}}
    a_i({\bf k}){1\over 2}
   \left[\Gamma(\nu)\left({\xi\over 2}\right)^{-\nu}\!\!\!\! 
   + \Gamma(-\nu)\left({\xi\over 2}\right)^{\nu}+\dots\right].
\label{a_i_expansion} 
\end{equation}
Combining $A_i^{(g)}$ and $A_i^{(p)}$ we have
\begin{eqnarray} 
A_i (x_0, {\bf x}) 
   = x_0^{d/2}\int {d^d k \over (2\pi)^d} e^{-i{\bf k}\cdot{\bf x}}
   &&\!\!\!\!\!\!\!\! 
\left[\, 
       \Gamma(\nu) \left( {\xi\over 2} \right)^{-\nu}\!\! 
     \left( {1\over 2} a_i({\bf k}) - i {k_i\over |{\bf k}|^2}ca_0({\bf k})
     \right)\right. \\
 +  && \!\!\!\!\!\!\!\!
\left.
    \Gamma(-\nu) \left( {\xi\over 2} \right)^{\nu}
 \left( {1\over 2} a_i({\bf k}) - i {k_i\over |{\bf k}|^2}\overline{c}a_0({\bf k})
 \right) +\dots.\,
\right].
\nonumber 
\end{eqnarray}                                                  

Now rewrite the classical action (\ref{final_action}) in terms of the following
$\epsilon$-boundary field,
\begin{eqnarray}
A_{\epsilon,i}({\bf x}) &=& \epsilon^{\lambda} A_i(\epsilon, {\bf x})\\
     &=& \int {d^d k \over (2\pi)^d} e^{-i{\bf k}\cdot{\bf x}} 
\left[\, 
       \Gamma(\nu) \left( { |{\bf k}|\over 2} \right)^{-\nu}\!\! 
     \left( {1\over 2} a_i({\bf k}) - i {k_i\over |{\bf k}|^2}ca_0({\bf k})
     \right)\right.  \nonumber\\  
 && \hskip2.06cm + \left. 
 \Gamma(-\nu) \left( { |{\bf k}|\over 2} \right)^{\nu}
 \left( {1\over 2} a_i({\bf k}) - i {k_i\over |{\bf k}|^2}\overline{c}a_0({\bf k})
 \right)\epsilon^{2\nu} +\dots.\,
\right],
\nonumber 
\end{eqnarray}           
where $\lambda$ is already defined to be $\nu - {d \over 2}.$ 
The last two terms of the action (\ref{final_action}) diverges as 
$\epsilon^{-2\nu}$
or $\epsilon^{-2\nu+2}$ as $\epsilon\to 0,$  and we ignore as usual.
Using (\ref{example}), the relevant part of the action is 
\begin{equation}
I=-{1\over 4}\lim_{\epsilon\to 0}\epsilon^{-2\nu}\int d^dx\;
  \epsilon \partial_\epsilon A_{\epsilon,i}({\bf x})^2.
\end{equation}
Consider the $\epsilon\to 0$ limit of this action.  
To clarify the meaning of the coefficients $a_i({\bf k})$ and $a_0({\bf k})$
we consider the following Fourier transformation,
\begin{equation}
\lim_{\epsilon\to 0} A_{\epsilon,i}({\bf x}) = \int {d^d k \over (2\pi)^d} 
e^{-i{\bf k}\cdot {\bf x}} \tilde{A}_i({\bf k}).
\end{equation} 
It is clear that the Fourier component is 
\begin{equation}
\tilde{A}_i({\bf k}) = 
 \Gamma(\nu)\left({ |{\bf k}| \over 2}\right)^{-\nu}
  \left( {1\over2}a_i({\bf k}) - i{k_i\over|{\bf k}|^2}ca_0({\bf k}) \right).
\end{equation}

To solve  $a_i({\bf k})$ and $a_0({\bf k})$ in terms of $d-$dimensional field
components
$\tilde{A}_i({\bf k})$ we use  (\ref{k_ia_i=0}),
\begin{eqnarray}
a_0({\bf k}) &=& {i\over c\Gamma(\nu)}
    \left({|{\bf k}|\over 2}\right)^\nu k_i\tilde{A}_i({\bf k}),\\
a_i({\bf k}) &=& {2\over \Gamma(\nu)}\left({|{\bf k}|\over 2}\right)^\nu
\left(\delta_{ij} - {k_i k_j \over |{\bf k}|^2}\right) \tilde{A}_j({\bf k}).
\end{eqnarray}
Substituting these two relations into (\ref{final_action})
we have 
\begin{equation}
I = -\nu {\Gamma(-\nu) \over \Gamma(\nu)} \int {d^dk \over (2\pi)^d}
    \tilde{A}_i({\bf k}) \left[ (\delta_{ij} - {k_i k_j \over |{\bf k}|^2} )
    +{\bar{c} \over c} {k_i k_j \over |{\bf k}|^2} \right] 
    \left({|{\bf k}| \over 2}\right)^{2\nu}\tilde{A}_j(-{\bf k}).
\end{equation}
From the specific forms (\ref{c}) it can be shown that
\begin{equation}
\delta_{ij} - {k_i k_j \over |{\bf k}|^2}
 +{\bar{c} \over c} {k_i k_j \over |{\bf k}|^2}
= \delta_{ij} + {2\nu \over 1- {d\over 2}-\nu} {k_i k_j \over |{\bf k}|^2}.
\end{equation}
To get the final action we use (A.1-A.3) of the appendix,
\begin{equation}
I =-c_{\Delta, d} \int d^d x d^d x'
   A_i({\bf x}) 
   \left( {\delta_{ij}\over |{\bf x} -{\bf x}'|^{2\Delta}}
- 2{ (x-x')_i (x-x')_j \over |{\bf x} -{\bf x}'|^{2\Delta+2} }\right)
   A_j({\bf x'}),
\end{equation}
where $\Delta=(\lambda+d).$ 
The proportional constant $c_{\Delta, d}$ which is related to the central
charge\cite{Witten's_path,central_charge} is given by
\begin{equation}
c_{\Delta, d} ={ \Delta(\Delta-{d\over 2})\Gamma(\Delta-1)
       \over\pi^{d\over 2} \Gamma(\Delta-{d\over 2})}.
\end{equation}
This result coincides exactly with that of \cite{Muck}.
The two point correlation function $\langle J_i({\bf x}) J_j({\bf x}')\rangle$ 
of the conformal current,which can be read off from the Maldacena's
proposal (\ref{proposal}--\ref{proposal_2}) is 
\begin{equation}
\langle J_i({\bf x}) J_j({\bf x}')\rangle = c_{\Delta, d}
\left({\delta_{ij}\over |{\bf x} -{\bf x}'|^{2\Delta}}
- 2{ (x-x')_i (x-x')_j \over |{\bf x} -{\bf x}'|^{2\Delta+2} } \right),
\end{equation}
showing that $\Delta$ is the corresponding conformal dimension.  
One may check the resulting $\Delta$ with the known value 
for the special case of $m=0.$
In this case the larger solution of (\ref{lambda}) is $\lambda = -1.$
This means the conformal dimension $\Delta$ is equal to $d-1$ which 
in fact agrees with the Witten's result.\cite{Witten_holography}
\section*{Acknowledgement}
The author would like to thank
A.~Petkou for his helpful comments on the various facets of 
the holographic projection.
This work is supported in part by NON DIRECTED RESEARCH FUND, 
Korea Research Foundation, 1997, and in part by the Basic Science Research
Institute Program of the Ministry of Education, Korea, BSRI-97-2436.

\section*{Appendix}
\renewcommand{\theequation}{A.\arabic{equation}}
\setcounter{equation}{0}

It is known\cite{Fourier} that for non-integer $\delta,$ 
\begin{equation}
\int {d^dk\over (2\pi)^d}  \, e^{i{\bf k}\cdot{\bf x}} |{\bf k}|^{\delta} = 
{C_{d,\delta}\over |{\bf x}|^{d+\delta}},
\end{equation}
where $C_{d,\delta}$ is a constant which is given by
\begin{equation}
C_{d,\delta}={2^\delta \over \pi^{d\over 2} }
 {\Gamma({d+\delta \over 2}) \over \Gamma(-{\delta\over 2})}.
\end{equation}
Differentiating this with respect to $x^{i}$ and $x^{j}$ we have following
relation,
\begin{equation}
\int {d^dk \over (2\pi)^d} \, e^{i{\bf k}\cdot{\bf x}} {k_i k_j \over |{\bf k}|^2} |{\bf k}|^{\delta}   
 = -{1\over\delta}\left\{ \delta_{ij} - (\delta+d) {x_i x_j \over |{\bf x}|^2}
  \right\} {C_{d,\delta}\over |{\bf x}|^{d+\delta}}.
\end{equation}
One may check the validity of this equation by contracting $ij-$indices and by
comparing with (A.1).
\vskip1cm

\end{document}